\documentstyle[12pt]{article}
\catcode`\@=11
\def\makepreprititle{\par
\begingroup
\def\thefootnote{\fnsymbol{footnote}}
\def\
@makefnmark{\hbox 
to 0pt{$^{\@thefnmark}$\hss}} 
\if@twocolumn 
\twocolumn[\@makepreprititle] 
\else \newpage
\global\@topnum\z@ 
\@makepreprititle \fi\thispagestyle{empty}\@thanks
\endgroup
\setcounter{footnote}{0}
\let\makepreprititle\relax
\let\@makepreprititle\relax
\gdef\@thanks{}\gdef\@author{}\gdef\@title{}
\gdef\@preprintnumber{}\gdef\@preprintdate{}\gdef\subtitle{}
\let\thanks\relax}
\def\preprintnumber#1{\gdef\@preprintnumber{#1}}
\def\preprintdate#1{\gdef\@preprintdate{#1}}
\def\subtitle#1{\gdef\@subtitle{#1}}
\def\@makepreprititle{\newpage
{\def\baselinestretch{1}
\begin{flushright} \@preprintnumber \par
\@preprintdate \end{flushright} } \par
\begin{center}
\vskip 1.5em
{\LARGE \@title \par} \vskip 2.5em 
{\large \lineskip .5em
\begin{tabular}[t]{c}\@author 
\end{tabular}\par}
\vskip 1em {\large \@date} \end{center}
\par
\vfil} 
\date{\sl Department of Physics, Tohoku University\\Sendai, 980 Japan}
\preprintdate{~}
\preprintnumber{~}
\subtitle{}
\def\endabstract{\if@twocolumn\else\endquotation\fi}

\catcode`\@=12

\def\lsim{\mathrel{\mathpalette\vereq<}}
\def\gsim{\mathrel{\mathpalette\vereq>}}
\catcode`\@=11
\def\vereq#1#2{\lower3pt\vbox{\baselineskip1.5pt \lineskip1.5pt
\ialign{$\m@th#1\hfill##\hfil$\crcr#2\crcr\sim\crcr}}}
\catcode`\@=12
\renewcommand{\thefootnote}{\fnsymbol{footnote}}
\preprintdate{June 1994}
\preprintnumber{LBL-35774\\RU-94-51}
\title{Upper bound on Hot Dark Matter Density\\from $SO(10)$ Yukawa
Unification\thanks{This work was supported by the Director, Office of Energy 
Research, Office of High Energy and Nuclear Physics, Division of High
Energy Physics of the U.S. Department of Energy under Contract
DE-AC03-76SF00098, and by NSF grants PHY-91-21039 and PHY-89-04035.}}
\author{Andrea Brignole,$^1$\thanks{Supported by a fellowship of
Istituto Nazionale di Fisica Nucleare, Sezione di Padova, Padua, Italy.}
Hitoshi Murayama$^1$\thanks{On leave of
absence from {\sl Department of Physics, Tohoku University, Sendai, 980
Japan.}} and Riccardo Rattazzi$^2$}
\date{\sl $^1$ Theoretical Physics Group, Lawrence Berkeley Laboratory\\
University of California, Berkeley, CA 94720\\
$^2$ Department of Physics and Astronomy, Rutgers University\\
Piscataway, NJ 08855}
\begin{document}
\thispagestyle{empty}
\makepreprititle

\begin{abstract}

We study low-energy consequences of supersymmetric $SO(10)$ models
with Yukawa unification $h_t = h_N$ and $h_b = h_\tau$. 
We find that it is difficult to reproduce the observed $m_b/m_\tau$ ratio
when the third-generation right-handed neutrino is at an intermediate
scale,
especially for small $\tan \beta$.
We obtain a conservative lower bound on the mass of the right-handed neutrino 
$M_N > 6 \times 10^{13}$~GeV for $\tan \beta < 10$.
This bound translates into an upper bound on the $\tau$-neutrino mass, and
therefore on its contribution to the hot dark matter density of the present 
universe, $\Omega_\nu h^2 < 0.004$.
Our analysis is based on the full two-loop 
renormalization group equations with one-loop threshold effects. 
However, we also point out that physics above the
GUT-scale could modify the Yukawa unification condition $h_b = h_\tau$
for $\tan \beta 
\lsim 10$. This might affect the prediction of $m_b/m_\tau$ and the
constraint on $M_N$.
\end{abstract}

\newpage
\renewcommand{\thepage}{\roman{page}}
\setcounter{page}{2}
\mbox{ }

\vskip 1in

\begin{center}
{\bf Disclaimer}
\end{center}

\vskip .2in

\begin{scriptsize}
\begin{quotation}
This document was prepared as an account of work sponsored by the United
States Government. While this document is believed to contain correct
 information, neither the United States Government nor any agency
thereof, nor The Regents of the University of California, nor any of their
employees, makes any warranty, express or implied, or assumes any legal
liability or responsibility for the accuracy, completeness, or usefulness
of any information, apparatus, product, or process disclosed, or represents
that its use would not infringe privately owned rights.  Reference herein
to any specific commercial products process, or service by its trade name,
trademark, manufacturer, or otherwise, does not necessarily constitute or
imply its endorsement, recommendation, or favoring by the United States
Government or any agency thereof, or The Regents of the University of
California.  The views and opinions of authors expressed herein do not
necessarily state or reflect those of the United States Government or any
agency thereof of The Regents of the University of California and shall
not be used for advertising or product endorsement purposes.
\end{quotation}
\end{scriptsize}

\vskip 2in

\begin{center}
\begin{small}
{\it Lawrence Berkeley Laboratory is an equal opportunity employer.}
\end{small}
\end{center}

\newpage
\renewcommand{\thepage}{\arabic{page}}
\setcounter{page}{1}

{\bf 1.} Grand unification \cite{GG} has been one of the leading guiding
principles to build models of fundamental forces in nature. The 
introduction of a simple gauge group
not only allows  an elegant description of the electromagnetic, weak and
strong forces, but also gives powerful insight into the structure of
Yukawa couplings \cite{Chanowitz}.
Supersymmetry, on the other hand, has provided a natural framework to
solve the
gauge hierarchy problem. 
On top of these theoretical ideas, there also appeared an
experimental indication supporting supersymmetry combined with grand
unification.
Experiments as LEP \cite{LEP} and SLC \cite{SLC} have provided a very
precise 
determination of the weak mixing angle. The resulting value has
ruled out the minimal grand unified theories (GUTs), while showing a
remarkable agreement 
with their supersymmetric versions based on the minimal particle content 
below the GUT scale $M_G$ \cite{costa,unification}.
(The absence of particle
thresholds between the weak and GUT scales is often referred to as
``grand desert assumption''.) In the same class of theories,
the partial unification of Yukawa interactions ($h_b=h_\tau$) has also
proven successful \cite{Arason,BBO,Carena}, while in the non-SUSY models it
has not. These
facts have strengthened the motivation for studying more detailed 
consequences of SUSY-GUTs with the grand desert. Recent work includes
more theoretical efforts on GUT model building as well as more
phenomenological studies such as those on cosmic neutralino abundance,
proton decay and collider signatures.

On the other hand, there are indications from astrophysics and cosmology
that an intermediate scale might exist in the middle of the proposed grand
desert. One indication comes from the observed deficit in the solar neutrino
flux \cite{Bahcall}, which favors 
neutrino oscillations {\it \`a la}\/ Mikheev--Smirnov--Wolfenstein (MSW)
\cite{MSW} with non-vanishing neutrino masses of ${\cal O}(10^{-3})$~eV
\cite{GALLEX2}. If we
regard the neutrino oscillation as being of $\nu_e$--$\nu_\mu$
type,
then the tiny masses required for the MSW effect suggest the existence of
right-handed neutrinos having an intermediate mass of
$10^{10}$--$10^{13}$~GeV and acting as the 
source of a seesaw mechanism \cite{seesaw}.\footnote{We are assuming
$SO(10)$-like mass 
relations, where the neutrino Dirac masses are of the order of the
up-type quark masses.} Another indication comes
from the density fluctuations at large
scales observed by COBE \cite{COBE}, which have set the normalization of the
Harrison--Zeldovich spectrum. The result, if one assumes that dark matter
is only of the cold type, is in a moderate contradiction with
the observations at smaller scales \cite{Maddox,Davis}. 
It has been suggested that the
existence of both cold and hot dark matter components solves this
problem
\cite{Primack}. The only particle physics candidate for hot dark matter
is a light neutrino with a mass around 10~eV. If we regard this neutrino as
$\nu_\tau$, then, again, a right-handed neutrino at an intermediate scale
$\sim 10^{12}$~GeV is favoured.
Moreover, it was also shown that scalar components of the right-handed
neutrinos can play interesting cosmological roles in baryogenesis
\cite{MY} and inflation \cite{MSYY}. 

Therefore, it is important to study the phenomenology of SUSY
GUTs in which right-handed neutrinos are the only thresholds in the
grand desert.
To our knowledge, few studies of the
low-energy phenomenology have been done within this
context (see, {\it e.g.}, \cite{Masiero,Moroi}).
Obviously, gauge coupling constant
unification is not significantly affected because the right-handed neutrinos
are gauge singlets. 

This letter studies the unification of the Yukawa couplings of the third
family in SUSY GUTs
with right-handed neutrinos at an intermediate scale. We base our
analysis on $SO(10)$ models with Yukawa unification
 $h_t = h_N$ and $h_b=h_\tau$.
We find that an intermediate scale right-handed neutrino has a critical
influence on the predicted value of $m_b/m_\tau$.
Consistency with the experimental value of $m_b/m_\tau$ implies
 a lower bound on
the mass of the right-handed neutrino. This constraint translates into an
upper bound on the $\tau$ neutrino contribution $\Omega_\nu$ to the
hot dark matter density in the present universe. Cosmologically
interesting values of $\Omega_\nu$ are strongly disfavored
for small values of $\tan \beta$.\footnote{There is a claim
\cite{Polonsky} that
$b$--$\tau$ Yukawa unification combined with the 
value of the top quark mass recently suggested by CDF excludes
the region $1.5 \leq \tan \beta \leq30$. 
 This claim, however, depends
crucially on $\alpha_s (m_Z) \simeq 0.125$ and $m_t \lsim 174$~GeV,
for which the experimental support is not yet established.
Thus, we take $\tan \beta$ as a free parameter in this letter.}

{\bf 2.} We first describe the theoretical framework of our study.
We assume a simple $SO(10)$ GUT scenario 
where only {\it small}\/ irreducible representations (irreps) like ${\bf
10}$, ${\bf 16}$, $\overline{\bf 16}$, ${\bf 45}$, ${\bf 54}$ and
possibly singlets are present,\footnote{Notice that {\it large}\/ irreps
drive the theory out of the perturbative regime; for example, if 
a {\bf 126} is present, the $SO(10)$ gauge coupling constant blows up
below $< 8 
M_G$ even with the minimum particle content ${\bf 45}+{\bf 54}+{\bf
126}+\overline{\bf 126}+3\times {\bf 16}$. 
It is also interesting to recall that, in superstring models, the above
mentioned {\it small}\/ irreps appear already at Kac-Moody level $k=2$,
whereas {\bf 126} can appear only for $5\leq k \leq 7$ \cite{FIQ}.
Phenomenologically, the presence of a ${\bf 126}$ may also give an unacceptable
relation $h_\tau = 3 h_b$ \cite{HRS}. Finally, it is noteworthy that 
the absence of ${\bf 126}$
and R-odd gauge singlets forbids  right-handed neutrino masses at the
renormalizable level, thus providing  a dimensional argument for
their being considerably smaller than $M_G$.} and where
matter fields come only in ${\bf 16}$'s.
As for the Yukawa interactions, the large value of the top Yukawa coupling
suggests that it originates from a renormalizable coupling.  With our
assumptions this can only be $h_G {\bf 16_3 10_H 16_3}$.
(Even a ${\bf 120}$ would not couple because of antisymmetry.)
In general, the two light doublets $H_{u,d}$ of the 
Minimal Supersymmetric Standard Model (MSSM)
contain only fractions of those in ${\bf 10_H}$, due to mixing
with doublets in other irreps. In our case these  can be 
${\bf 10'_H}$'s, ${\bf 16_H}$'s and ${\bf \overline{16}_H}$'s,  the
latter ones being necessarily present in order to reduce the rank 
of the gauge group. The mixing strengths are in general
different for $H_u$ and $H_d$ because $SU(2)_R$ is broken at $M_G$.
Then the GUT
boundary condition on the Yukawa couplings is
$h_t = h_N=s_uh_G, \,\,\, h_b = h_\tau=s_d h_G$, 
where $s_{u,d}$ are the Higgs mixing angles, and in general $s_u\not = s_d$.
Here the meaning of $h_t$, $h_N$, $h_b$, and $h_\tau$ is the obvious one.
The particular case $s_u=s_d$  has been 
studied before \cite{ALS,HRS}. 
Actually the boundary condition 
\begin{equation}
h_t = h_N, \,\,\, h_b = h_\tau  \label{boundary}
\end{equation}
is just a result of the underlying Pati-Salam $SU(4)$ symmetry 
\cite{PS}, so that one may imagine scenarios, maybe in strings,
where eq.~(\ref{boundary}) 
holds even without a unified simple gauge group.  There are also more
exotic scenarios with mixings between ${\bf 16}$ and ${\bf 10}$ matter fields
\cite{splitting}, while eq.~(\ref{boundary}) keeps holding.
Moreover, there is a class of models where $h_b$ and $h_\tau$ originate
solely from higher-dimension operators, while predicting the same
relation \cite{KS}. We thus conclude that there is a wide class of models
($SO(10)$ and beyond) which lead to the boundary condition
eq.~(\ref{boundary}).

{\bf 3.} In order to illustrate the low-energy consequences of $b$--$\tau$ 
Yukawa unification when a right-handed neutrino is present,
we first consider a simplified picture based on one-loop 
renormalization group equations (RGE), without threshold
corrections. This also allows for a clearer comparison
with the pure MSSM case, {\it i.e.} the case without 
right-handed neutrinos below the GUT scale.
The RGE for the individual gauge and Yukawa couplings can be
derived from the general formulae in Ref. \cite{BJ}, 
both at one-loop order (needed here) and at two-loop order 
(to be used below), and we do not write them down explicitly. 
Here we rather focus on the evolution of
the ratio $R(\mu)\equiv h_b(\mu)/h_\tau(\mu)$ from the GUT-scale 
value $R(M_G)=1$ to the low-energy value $R(m_Z)$. The latter 
should be compared with the experimental value of $m_b/m_{\tau}$
scaled up to $m_Z$, which we denote by $R^{exp}(m_Z)$. 
We first recall the situation in the pure MSSM case \cite{IL}.
The RGE for $R$ reads 
\begin{equation}
\label{mssm}
\frac{{\rm d} R}{{\rm d}t}
	= \frac{R}{16 \pi^2} 
	\left[ h_t^2 + 3 (h_b^2 - h_\tau^2)
		- (\frac{16}{3}g_3^2-\frac{4}{3} g_1^2)  \right],
\end{equation}
where $t \equiv \log {\mu}$. If all the Yukawa coupling constants 
are neglected, one obtains for $R(m_Z)$ the expression
\begin{equation}
\label{gauge}
 R_g \equiv \left( \frac{\alpha_3(m_Z)}{\alpha_3(M_G)} 
\right)^{\frac{8}{9}}
\left( \frac{\alpha_1(m_Z)}{\alpha_1(M_G)} \right)^{\frac{10}{99}},
\end{equation}
which is numerically larger than $R^{exp} (m_Z)$ by 30--80~\%. 
The Yukawa contributions in eq.~(\ref{mssm}) are essential in reducing
$R(m_Z)$ from 
its `pure gauge' value $R_g$ to the observed value  $R^{exp}(m_Z)$.
Actually, one could formally obtain arbitrarily small values for $R(m_Z)$ 
by taking large enough $h_t$ at the  GUT-scale \cite{IL}.
In practice, initial values of ${\cal O}(1)$ are required to reproduce
the correct  $R^{exp}(m_Z)$.
On the other hand, if a right-handed neutrino with mass $M_N < M_G$ 
is present,  the MSSM eq.(\ref{mssm}) applies only for  $m_Z < \mu < M_N$.
For $M_N < \mu < M_G$ the theory is the MSSM+N (in an obvious notation)
and the RGE has a slight (but crucial) modification: 
\begin{equation}
\label{mssmn}
\frac{{\rm d} R}{{\rm d}t}
	= \frac{R}{16 \pi^2} 
	\left[ (h_t^2-h_N^2) + 3 (h_b^2 - h_\tau^2)
		- (\frac{16}{3}g_3^2-\frac{4}{3} g_1^2)  \right].
\end{equation}
The essential difference with the pure MSSM case is a manifest 
cancellation of {\em all} the Yukawa contributions, at least
close to the GUT-scale, due to the boundary conditions 
$h_t = h_N$ and $h_b = h_\tau$. 
As a consequence, even if one formally allows for arbitrarily 
large values of the Yukawa couplings at the GUT-scale, the low-energy 
value $R(m_Z)$ cannot be made arbitrarily small, in contrast 
to the pure MSSM case. In particular, it may be
impossible to reach the experimental value $R^{exp}(m_Z)$. 

To make this point more explicit, it is useful to integrate 
formally eqs.(\ref{mssm}) and (\ref{mssmn}),  and write 
$R(m_Z)$ as 
\begin{equation}
R(m_Z) = R_g  \cdot e^{-Y} ,
\end{equation}
where
\begin{equation}
\label{yyy}
Y = \frac{1}{16\pi^2}\left[ \int_{t_Z}^{t_N} h_t^2(t) {\rm d}t 
+ \int_{t_N}^{t_G} (h_t^2(t)-h_N^2(t))  {\rm d}t
+ 3 \int_{t_Z}^{t_G} (h_b^2(t)-h_{\tau}^2(t))  {\rm d}t \right],
\end{equation}
with $t_{A} \equiv \log {M_{A}}$. Notice that $Y$ should be considered
as a function of the initial values of the Yukawa couplings and
the mass of the right-handed neutrino.
Since all the three terms in $Y$ turn out to be positive, 
$R(m_Z)$ is smaller than $R_g$ also in the present case.
The question is whether or not $Y$ can be large enough to reproduce 
the correct $R^{exp}(m_Z)$.
Indeed the cancellation effect mentioned above limits 
the size of $Y$, even for large initial Yukawa couplings.
For a given $M_N<M_G$, $Y$ has a (finite) upper bound $Y_{max}(M_N)$,
formally obtained by taking infinite Yukawa couplings 
at $M_G$. Since $Y_{max}(M_N)$ becomes smaller for 
smaller values of $M_N$, it may be impossible to get $R(m_Z)$ as low
as $R^{exp}(m_Z)$ when $M_N$ is lighter than a
certain critical value. This is why a lower bound on $M_N$ appears.
Actually, $Y_{max}(M_N)$ crucially depends on $\tan\beta$. 
For fixed $M_N$, $Y_{max}(M_N)$ is smaller when $\tan\beta$ 
is small, because $h_b$ and $h_\tau$  are negligible and 
the last term in $Y$ is missing. In fact this is the
situation where stronger constraints on $M_N$ will be found.
On the contrary, when  $\tan\beta$ is very large, the
last term in $Y$ plays a significant role (notice also
the factor 3). Then, even if $M_N$ is light, $Y$ can be
made large enough as to reduce the value of $R$ appropriately.

The above qualitative discussion is supported not only by
a numerical analysis, but also by an analytical argument 
which we sketch below. The argument establishes the existence of 
an upper bound on $Y$ as a function of $M_N$ independent
of the initial conditions on the Yukawa couplings. Even
if one cannot solve exactly the coupled system of RGE,
exact inequalities can be found that lead to the final result.
We consider here the most interesting case where $h_b$ and $h_\tau$ are 
negligible. Then we can write $Y=Y_1+Y_2$, where $Y_1$ and $Y_2$
are the first and second terms in eq.~(\ref{yyy}).
We denote by $h_0(t)$ the solution of the RGE for $h_t(t)$ in
the pure MSSM case with boundary condition 
$h_0(M_G)\rightarrow \infty$. Its explicit expression is \cite{IL}
$h_0^2(t) = 4 \pi^2 (- F'(t)/3F(t))$, where
$F(t) \equiv \int_t^{t_G} {\rm d} t'
\left( {\alpha_3(M_G)}/{\alpha_3(t')} \right)^{-\frac{16}{9}}
\left( {\alpha_2(M_G)}/{\alpha_2(t')} \right)^{3}
\left( {\alpha_1(M_G)}/{\alpha_1(t')} \right)^{\frac{13}{99}}$.
A study on the RGE of $h_t(t)$ for $t<M_N$ and $t>M_N$, with
a given boundary condition $h_t(M_G)$, leads easily to the
absolute upper bound $h_t(t) < h_0(t)$, valid for any $t$
and independent of $h_t(M_G)$. 
This gives immediately an upper bound on $Y_1$, 
\begin{equation}
\label{y1b}
Y_1 < Y_1^*(M_N) \equiv \frac{1}{12} \log\frac{F(m_Z)}{F(M_N)}
\end{equation}
where we have replaced for convenience the arguments $t_A$ with $M_A$.
A simple bound on $Y_2$ can 
be derived {\it e.g.} by studying the RGE for the ratio 
$\tilde{R}(t) \equiv h_t(t)/h_N(t)$, which reads 
\begin{equation}
\label{rprime}
\frac{1}{\tilde{R}}\frac{{\rm d}\tilde{R}}{{\rm d}t}
	= \frac{1}{16 \pi^2} 
	\left[ 3(h_t^2-h_N^2) 
	- (\frac{16}{3}g_3^2+\frac{4}{15} g_1^2)  \right].
\end{equation}
If one formally integrates the above equation term by term 
between $t_N$ and $t_G$, the right hand side contains $Y_2$ 
and easily integrable gauge terms. The left hand side gives 
just $-\log \tilde{R}(t_N)$, which is negative (one can easily prove 
that $\tilde{R}(t)>1$ from a different form of eq.(\ref{rprime})). 
Therefore one obtains an absolute bound
\begin{equation}
\label{y2b}
Y_2 < Y_2^*(M_N) \equiv \frac{8}{27}\log \frac{\alpha_3(M_N)}{\alpha_3(M_G)} 
- \frac{2}{297}\log \frac{\alpha_1(M_N)}{\alpha_1(M_G)} ,
\end{equation}
Notice that the bounds (\ref{y1b}) and (\ref{y2b})
are independent of the initial values $h_t(M_G)=h_N(M_G)$,
so they hold even in the formal limit $h_t(M_G)\rightarrow \infty$.
In conclusion, for a given $M_N < M_G$ we obtain explicitly
an absolute upper bound $Y_{max}^*(M_N)$ on $Y$ which is 
a monotonically increasing 
function of $M_N$\footnote{ Although $Y_{max}(M_N)$ is always 
monotonically increasing, 
the same applies to the analytic bound $Y_{max}^*(M_N)$ 
for $M_N \gsim 10^{10}$ GeV only. However, this is just the region 
of interest. We add that stronger analytic bounds than
the simple one described here can be found.}
\begin{equation}
\label{anbound}
Y \leq Y_{max} (M_N) < Y_{max}^*(M_N) =Y_1^*(M_N)+Y_2^*(M_N) .
\end{equation}
This in turn gives an absolute lower bound on $R(m_Z)$
\begin{equation}
\label{Rbound}
R(m_Z) > R_g \cdot e^{-Y_{max}^*(M_N)} = R_g \cdot
\left(\frac{F(m_Z)}{F(M_N)}\right)^{-\frac{1}{12}}
\left( \frac{\alpha_3(M_N)}{\alpha_3(M_G)} \right)^{-\frac{8}{27}}
\left( \frac{\alpha_1(M_N)}{\alpha_1(M_G)} \right)^{\frac{2}{297}},
\end{equation}
proving that one cannot make $R(m_Z)$ arbitrarily small even
in the formal limit $h_t(M_G)\rightarrow \infty$. This bound
can be compared with $R^{exp}(m_Z)$, so that a lower bound 
on $M_N$ can be inferred.
 
{\bf 4.} We have shown an analytic bound on $R$ in which $h_N$ was allowed to
go to infinity at $M_G$. In practice $h_N$ will not be taken larger than
${\cal O}(1)$. Then the effect of a right handed neutrino with mass
around $10^{13}$~GeV can be roughly estimated by treating it as a
threshold correction to $R$ at $M_G$ \cite{HRS}. This gives $R \simeq 1 +
\frac{h_N^2}{16\pi^2} \log(M_G/M_N)$, which amounts to an ${\cal O}(10\%)$
increase in $R$. This 
$10\%$ is a critical amount when comparing to the experimental value, 
since already in the pure MSSM case 
the predicted $R$ lies in the upper part of the allowed range.
Nonetheless it is clear that all other effects  of the same order may
have a huge impact 
on the bound on $M_N$. This is because the correction
to $R$ from the right-handed neutrino depends roughly logarithmically on
$M_N$. Then, in order 
to make our analysis more accurate, we have to discuss SUSY and
GUT thresholds and use two-loop RGE. In fact the effects of these, and
especially the first two,
can possibly pile up to ${\cal O}(10\%)$. We also attempt an
estimate of the possible effects from physics above $M_G$, even
though we are aware that our understanding of these can only be  qualitative.
Finally,  by varying the QCD coupling $\alpha_s(m_Z)$ within its
experimental range we get an ${\cal O}(10$--$20\%)$ effect, which we
treat with
the proper attention. Our reference value of $\alpha_s(m_Z)$
is the one from the $Z$-boson hadronic width
$\alpha_s(m_Z)=0.124\pm 0.007$ \cite{miquel}, and we allow a 2~$\sigma$ range.

To obtain quantitative constraints on the mass of the right-handed
neutrino, we perform a numerical analysis based on two-loop 
RGE with one-loop threshold corrections. 
In order to be able to discuss the dependence on 
$\alpha_s (m_Z)$ \cite{BBO,HRS}, we take the following procedure. 
We {\it define}\/ $M_G$ as the scale where $\alpha_1$ and $\alpha_2$ meet,
and take $h_t=h_N$, $h_b=h_\tau$ at the same scale.\footnote{Numerically,
$M_G$ turns out to be around $2 \cdot 10^{16}$ GeV. GUT threshold
corrections on the Yukawa unification conditions will be discussed
later.} On the other hand,
we take $\alpha_s (m_Z)$ as an independent parameter, and allow
$\alpha_3 \neq \alpha_1 = \alpha_2$ at $M_G$. This difference
could be accounted for by GUT threshold effects. 
The gauge coupling constants in the MSSM at $m_Z$ are 
$\alpha_1^{-1} (m_Z) = 59.1 + \frac{1}{2\pi}
\left(\frac{12}{5} \log \frac{m_{SUSY}}{m_Z} + \frac{1}{10} \log
\frac{m_A}{m_Z} \right)$, $\alpha_2^{-1} (m_Z) = 29.4 + \frac{1}{2\pi} ( 4
\log \frac{m_{SUSY}}{m_Z} + \frac{1}{6} \log \frac{m_A}{m_Z} )$, and
$\alpha_3^{-1} (m_Z) = \alpha_s^{-1} (m_Z) + \frac{4}{2\pi} \log
\frac{m_{SUSY}}{m_Z}$. Here the logarithms take care of
the matching between the SM and the MSSM; we take the additional Higgs
doublet at $m_A$ and the rest of the SUSY particles at $m_{SUSY}$.
(The non-logarithmic parts and the translation from  $\overline {\rm MS}$ to
$\overline {\rm DR}$ are numerically small and have been neglected.) 
We also match the couplings in the MSSM+N to those in the MSSM at
$\mu=M_N$,\footnote{We define 
the right-handed neutrino mass $M_N$ by $M_N = \overline{M}_N 
(M_N)$, where $\overline{M}_N (\mu)$ is the $\overline{\rm
DR}$ running mass.}
\begin{eqnarray}
\label{hmatch}
\begin{array}{r@{=}lr@{=}l}
\left. h_b(M_N)\right|_{\rm MSSM} & \left. h_b(M_N)\right|_{\rm MSSM+N},&
\left. h_\tau(M_N)\right|_{\rm MSSM} &
	\left. h_\tau(M_N)\right|_{\rm MSSM+N} (1 - \varepsilon),\\
\left. h_t(M_N)\right|_{\rm MSSM} &
	\left. h_t(M_N)\right|_{\rm MSSM+N} (1 - \varepsilon),&
\left. h_N(M_N)\right|_{\rm MSSM} &
	\left. h_N(M_N)\right|_{\rm MSSM+N} (1 - 2 \varepsilon),
\end{array}
\end{eqnarray}
where $\varepsilon = \frac{h_N^2(M_N)}{32\pi^2}$. $h_N(\mu)$ in the MSSM
is defined in 
such a way that the coefficient of the dimension-five operator $(L
H_u)^2$ is $h_N^2(\mu)/M_N$. 

For any fixed values of $(h_t, h_b)$ at $M_G$, the actual
calculation is done as follows. We use the two-loop RGE of the MSSM and
the MSSM+N. 
By using an iterative procedure, we find the values of $M_G$, 
$\alpha_1(M_G)$ and $\alpha_3(M_G)$ which reproduce 
the values of $\alpha_1 (m_Z)$, $\alpha_2 (m_Z)$ 
and $\alpha_3 (m_Z)$ defined above. 
At the same time we obtain the values of the Yukawa couplings at $m_Z$.
The resulting  $b$--$\tau$ Yukawa ratio 
$\displaystyle \left. R(m_Z)\right|_{\rm MSSM}$ is
then matched to the ratio of the corresponding masses in the broken
electroweak theory  $R(m_Z) = m_b(m_Z)/m_\tau (m_Z)$,
\begin{equation}
R(m_Z) = (1 + k_b - k_\tau + f_R) \left. R(m_Z)\right|_{\rm MSSM},
\end{equation}
where
$k_b$, $k_\tau$ are the threshold corrections due to the SUSY
particles, while $f_R$ is that from the additional Higgs
doublet in the MSSM. The exact expressions are found in
Ref.~\cite{HRS}. 
In the following we will only focus on the logarithmic terms in 
$k_{b,\tau}$. We will discuss the potentially large
non-logarithmic ones ($k'_b$ in Ref.~\cite{HRS}) at the end of the letter.
Note that also $\left. R(m_Z)\right|_{\rm MSSM}$ depends implicitly
on $m_{SUSY}$ and $m_A$ due to the threshold dependence of the MSSM gauge
coupling constants $\alpha_i^{-1} (m_Z)$ (see above).
For instance, in the small $\tan \beta$ region,  the 
dependence of $R$ on the SUSY particle masses turns out to be 
approximately\footnote{
In a more general case where the SUSY spectrum is nondegenerate,
the dependence of $R$ on the SUSY particle masses can again
be summarized by an effective $m_{SUSY}$. It turns out that
the dependence of $R$ on colorless particle masses is very weak,
and $m_{SUSY}$ can be interpreted as a geometric average
of squark and gluino masses. After considering generic
splittings in the SUSY spectrum, we found that the value of $R$ varied 
only between two extreme cases, $m_{SUSY} = 1$~TeV, $m_A = m_Z$, 
which gives the smallest $R$, and $m_{SUSY}=m_Z$, $m_A = 1$~TeV 
which gives the largest $R$. We stress that the approximate formula
eq.~(\ref{approx}) cannot be applied for $m_A \lsim m_Z$.}
\begin{equation}
\frac{\delta R}{R} \simeq \frac{1}{2\pi} \left(
- 0.1 \, \log\frac{m_{SUSY}}{m_Z} + 0.1 \, \log\frac{m_A}{m_Z} \right).
\label{approx}
\end{equation}
The threshold corrections
roughly cancel when $m_{SUSY} = m_A$, which is consistent with the
results in Ref.~\cite{BBO}. In the final step, 
the predicted $R(m_Z)$ is compared to $ R^{exp}(m_Z)$.
Notice that also $R^{exp}(m_Z)$ has an important dependence on $\alpha_s (m_Z)$,
since it is obtained by scaling the $b$--$\tau$ mass ratio up to
$m_Z$.  GUT-scale threshold corrections could also affect the boundary 
condition in eq.~(\ref{boundary}). Though we neglect
such corrections in the following numerical analysis, we will discuss
their possible effects later.

We first illustrate the dependence of $R$ on $h_t (M_G)$ for the
region $\tan \beta \lsim 10$, where both $h_b$, 
$h_\tau$ get renormalized homogeneously and the result depends very weakly on
$\tan \beta$. 
We show curves for different values of
$M_N$ in Fig.~1, taking  $m_{SUSY} = m_A = m_Z$,
$h_b = h_\tau = 0.01$ at $M_G$, and $\alpha_s (m_Z) = 0.11$ (Fig.~1a)
or 0.12 (Fig.~1b). 
For comparison, we also show the values of $R^{exp}(m_Z)$ corresponding 
to the $\overline{\rm MS}$ mass $m_b(m_b) = 3.9$, $4.15$, and $4.4$~GeV.
This is the range which is obtained from 
QCD sum rules \cite{GL,HRS}.\footnote{The uncertainty on $m_b (m_b)$ is
dominated by our lack of knowledge on the ${\cal O}(\alpha_s^2)$
corrections to QCD sum rules. See Ref.~\cite{HRS}.}
First of all, it is clear from the figures that it becomes harder to
reconcile $R(m_Z)$ with $R^{exp} (m_Z)$ for larger $\alpha_s
(m_Z)$ and lower $M_N$. For $\alpha_s (m_Z) = 0.12$, one 
needs very large $h_t (M_G) \gsim 3$ even for $M_N = M_G$.\footnote{For
$\alpha_s (m_Z) \gsim 0.12$ and $h_t (M_G) < 2$, the predicted value of
$R(m_Z)$ is larger than the experimental upper bound. Consistency could
be possibly restored by allowing ${\cal O}(5\%)$ GUT- or SUSY-scale
threshold corrections. See discussion below eq.~(19).}
Moreover, even if increasing $h_t(M_G)$ makes $R(m_Z)$ decrease,
the suppression effect becomes weaker as we lower $M_N$. Note also that
the value of  $R(m_Z)$ depends only weakly on $h_t(M_G)$
for $h_t (M_G) \gsim 2$, especially for $M_N < 10^{14}$~GeV. Such a
`fixed point' 
behaviour and the dependence on $M_N$ were expected on the basis 
of the analytic discussion above  (see e.g. eqs.(\ref{anbound}-\ref{Rbound})).
As we lower $M_N$, we can check whether the curves reach the region of
$R^{exp}(m_Z)$ and then infer a lower bound on $M_N$.
In the following, when scanning the space of 
$(h_t(M_G), h_b(M_G))$, we will 
only take for definiteness
$h_{t,b}(M_G) < 2$. This reference value is motivated both by
the above observation and by perturbativity reasons (see also
point 2 below).

Next we show lower bounds on $M_N$ as functions of $\tan \beta$ in
Fig.~2. 
In order to obtain conservative bounds we take the maximum
value of $m_b (m_b) = 4.4$~GeV.
For most of the $\tan \beta$ values, we obtain stringent
lower bounds on $M_N$. We show the cases of two representative SUSY
particle spectra 
({\it a}) $m_{SUSY} = 1$~TeV (conservative) and ({\it b}) 
$m_{SUSY} = m_Z$ (more stringent), while always keeping $m_A = m_Z$ 
(conservative). Curves are shown for three 
values of $\alpha_s (m_Z) = 0.110$ (solid), 0.117 (dotted), and 0.125
(dashed), where the lower ones correspond to case ({\it a}) and the upper ones
to ({\it b}). The values of $M_N$ {\it above}\/ the curves are
consistent with $b$--$\tau$ Yukawa unification. 
The constraint becomes stronger for larger values
of $\alpha_s (m_Z)$, smaller values of $m_{SUSY}$, and larger values of
$m_A$.
The curves do not extend to the region
$\tan \beta \gsim 58$ because of the constraint $h_b = h_\tau < 2$ at
$M_G$. 
When the curves reach $M_G$, it means that the lower values of $\tan
\beta$ are not consistent with $b$--$\tau$ Yukawa unification even when
$M_N = M_G$ (with $h_t (M_G) < 2$). 
Possible effects from the non-logarithmic SUSY threshold corrections and
physics above the GUT-scale will be discussed in points (1) and (2) below.

\setcounter{footnote}{0}

The lower bound on $M_N$ has a very interesting implication for
the value of $\Omega_\nu$, the $\nu_\tau$ contribution to the 
hot dark matter density of the present universe.
The mass of $\nu_\tau$ is related to $M_N$ and $h_N$ via the seesaw
formula,\footnote{Here we neglect SUSY- and weak-scale
threshold corrections on $m_{\nu_\tau}$, which are at most ${\cal
O}(5\%)$. Neglecting them is also justified because
they affect the constraint on $M_N$ only linearly.
In contrast, the analogous corrections to $R(m_Z)$ discussed above 
are larger and affect the constraint on $M_N$ exponentially.}
\begin{equation}
\label{seesaw}
m_{\nu_\tau} = \frac{h_N^2 v^2 \sin^2 \beta}{M_N},
\end{equation}
where $v=174$~GeV and $h_N$ in the effective theory below $M_N$
was defined after eq.(\ref{hmatch}).
On the other hand, the cosmic energy density of a light 
neutrino of mass $m_\nu \lsim 1$~MeV is simply proportional to $m_\nu$
\cite{KT}, 
\begin{equation}
\label{omnu}
\Omega_\nu h^2 \simeq \frac{m_\nu}{\rm 91.5~eV}.
\end{equation}
Here, $h$ is the normalized Hubble constant, $h =
H_0/(100$~km\,sec$^{-1}$\,Mpc$^{-1})$. 
Note that $h_N (m_Z)$ cannot be larger than an infrared fixed
point value,
which is $h_N (m_Z) \leq 0.8$ for the cosmologically interesting region $M_N
\lsim 10^{13}$~GeV. 
Using that value in eqs.~(\ref{seesaw}-\ref{omnu}), we obtain the 
maximum possible value of $\Omega_\nu h^2$ for a given $M_N$. 
By requiring that $\nu_\tau$ gives a certain contribution to
$\Omega$, one can put an {\it upper bound}\/ on $M_N$. We indicated
such bounds for $\Omega_\nu h^2 = 0.01$, 0.1, and 1 in Fig.~2
with dotted lines.
We find
that the cosmologically interesting range $\Omega_\nu h^2 \gsim 0.1$
survives only for
large $\tan \beta$, especially for larger values of $\alpha_s
(m_Z)$. For $\alpha_s (m_Z) \gsim 0.125$, we find solutions only for
very large $\tan \beta$,
and they completely disappear for $\alpha_s (m_Z) >
0.13$. In other words, this analysis puts an upper bound on the cosmic
hot dark matter density as a function of $\tan \beta$. For instance, we
obtain
\begin{equation}
\Omega_\nu h^2 < 0.004,
\end{equation}
for $\tan \beta < 10$ even with the most conservative parameters
$\alpha_s (m_Z) = 0.11$ and case ({\it a}).

{\bf 5.} We now discuss the GUT-scale threshold corrections on $h_t=h_N$,
$h_b=h_\tau$. They mainly come from the mass splittings
within the heavy gauge multiplet as far as there are no other sizable Yukawa
couplings of ${\bf 16}_3$ to other GUT-scale fields \cite{HRS}. For
instance, mass splittings of ${\cal O}(10)$ modify the boundary condition by
less than $5$~\%.\footnote{As explained above, our procedure
implicitly requires a GUT-scale threshold correction on $\alpha_3
(M_G)$ to reproduce $\alpha_s (m_Z)$; however, the required correction
amounts to at most 5~\% if we vary $\alpha_s (m_Z) = 0.11$--$0.14$ and
$m_{SUSY} = m_Z$--1~TeV. Though the
corrections on $\alpha_3$ and 
$h_b/h_\tau$ are not directly correlated, this value gives us a
rough idea of the magnitude of the GUT-scale threshold corrections.}
A 5~\% correction on $h_t/h_N$ leads to less than
1~\% change in the prediction of $R(m_Z)$ so that it will not change the
constraint at all. However a 5~\% threshold correction on $h_b/h_\tau$ results
in a 5~\% correction in $R(m_Z)$, leading to a two order
of magnitude change on $M_N$ (see Fig.~1) and one might imagine a
situation where 
the combined effects of $\alpha_s(m_Z)=0.11$, SUSY and GUT thresholds 
pile up to allow $M_N=10^{12}$ GeV, even at small
$\tan\beta$. Situations like this are however rather extreme. When 
$\alpha_s(m_Z)\simeq 0.12$  (close to the centre of its range) and SUSY
particles are light (as they should be), then GUT threshold corrections must
amount to 15\% to make $M_N=10^{12}$ GeV possible, which
requires ${\cal O}(1000)$ mass splittings in the gauge multiplet.
Relaxation of the maximum allowed $h_t(M_G)$ up to 3.3 \cite{BBO} could
also weaken the constraint on $\Omega_\nu h^2$, but no more than by a
factor of 7. 

\setcounter{footnote}{0}

Our bound on
$\Omega_\nu$ has been derived in the approximation in which family mixings
are neglected. Then one might wonder whether, or under what conditions, the
mixings can sizably affect the bound. In order to evade our bounds,
the full mass matrices should satisfy the following two requirements:
i) The interaction eigenstate $N_3\subset {\bf 16_3}$ with Yukawa
coupling $h_N= h_t$ has an ${\cal O}(1)$ overlap with a mass
eigenstate of mass $\sim M_G$. Then the effective $h_N$ appearing in
the RGE below $M_G$ is smaller than $h_t$, and the effect of
$h_t$ in the running of $R$ is no longer ``exactly'' compensated.
ii) There still is a left-handed neutrino which contributes to a sizable
portion of $\Omega$. 
In order to study the consequences of these two requirements, we consider 
the case of two families, which may represent the second and third ones.
The superpotential reads as $W =
\hat{h}_e^{ij} L_i e_j H_d + \hat{h}_N^{ij} L_i N_j H_u + \frac{1}{2}
\hat{M}_{ij} N_i N_j$ with $i,j=2,3$, where $L$, $e$ and $N$ are, 
respectively, lepton doublet, right-handed
charged lepton, and right-handed neutrino superfields. The smallness of
CKM angles and $SO(10)$ relations among Yukawa matrices suggest that both
$\hat{h}_e^{ij}$ and $\hat{h}_N^{ij}$ are hierarchical in the same
basis. By going 
to the basis where ${\hat h}_N$ is diagonal, we parametrize
\begin{eqnarray}
\label{matrices}
\hat{h}_N & = & \left(\matrix{h_1&0\cr 0&h_2\cr}\right )
= \left(\matrix{\epsilon&0\cr 0&1\cr}\right )h_N,
\\
\hat M & = & \left(\matrix{M_{22}&M_{23}\cr M_{23}&M_{33}\cr}\right),
\end{eqnarray}
where $M_{ij} \lsim {\cal O}(M_G)$ and we expect 
$\epsilon \sim m_c/m_t \sim 10^{-2}$. 
Requirement i) implies that at least one of the $M_{ij}$ should be
${\cal O}(M_G)$, but not $M_{22}$ alone.  
The light neutrino mass matrix can be written as
\begin{equation}
\label{lightn}
{\cal M}_\nu  =  \langle H_u\rangle^2 \left(\hat{h}_N {\hat M}^{-1}
\hat{h}_N^{\rm T}\right)= 
\frac{\langle H_u\rangle^2 h_N^2}{{\rm Det}(\hat M)} 
\left(\matrix{\epsilon^2 M_{33}&-\epsilon M_{23}\cr -\epsilon M_{23}&
M_{22}\cr}\right ).
\end{equation}
Requirement ii) is expressed as
$\left (m_1^2+m_2^2\right )^{1/2}\geq 
{v^2 \sin^2 \beta}/{(\delta \cdot M_G)}$,
where $m_{1,2}$ are the eigenvalues of ${\cal M}_\nu$, and we need
$\delta\lsim 10^{-3}$ for a neutrino mass in the eV range. Therefore one
needs 
\begin{equation}
\label{finetune}
|{\rm Det} \hat{M}| \lsim \delta \left(\epsilon^4 M_{33}^2 
	+ 2 \epsilon^2 M_{23}^2 + M_{22}^2 \right)^{1/2} M_G ,
\end{equation}
where we used $h_N \sim 1$.
The above inequality implies a hierarchy
in the  eigenvalues of $\hat M$. This leads   to strong constraints
on $\hat{M}$. If all the entries of $\hat{M}$
are ${\cal O}(M_G)$, we need a fine-tuning of ${\cal O}(\delta)$ to obtain
the small determinant in eq.~(\ref{finetune}).\footnote{The tuning 
implies that that the leading 
contribution to $\hat M$ is rotated by ${\cal O}(1)$, with  respect to
the leading one in $\hat h_{e,N}$. It may be interesting to ask 
whether this can result from a flavour symmetry.}
We can avoid such a fine-tuning only if $\hat{M}$ is hierarchical
in the same basis where $\hat{h}$ is diagonal with
$M_{22}\lsim\delta \epsilon^2 M_G$, 
$M_{23}\lsim\delta^{1/2} \epsilon M_G$ and $M_{33} \sim M_G$.
In this case 
$\hat M$ could be viewed
as a simple consequence of an abelian horizontal symmetry, but notice that
the expansion parameter $\delta ^{1/2}\epsilon$ would be rather small. 
Moreover, the hot dark matter candidate would be predominantly $\nu_\mu$,
and the MSW oscillation should occur between $\nu_e$ and $\nu_\tau$. 
This is a new possibility which may be worth of
further study. Nonetheless both the latter possibility and the
fine-tuned case mentioned above show rather extreme features, thus
enhancing the 
importance of the cases to which our analysis correctly applies. These
cases include the reasonable situation in which there is no hierarchy among
right handed neutrino masses, and they all
decouple from the theory essentially at a single scale.

{\bf 6.} Finally we point out two generic uncertainties in $b$--$\tau$ Yukawa
unification which might affect the predictivity on
$R$. These exist even in the pure MSSM case. Of course, such
uncertainties may also affect the bound on $M_N$.

\renewcommand{\labelenumi}{(\arabic{enumi})}
\begin{enumerate}
\item SUSY-scale non-logarithmic threshold corrections to $R$
for large $\tan\beta$ \cite{HRS,Hempfling}.  
Typically the largest corrections  appear in the $b$ mass
via two diagrams, one involving gluino propagation and the other
involving higgsinos. The resulting correction to $m_b$ can 
be written as
\begin{equation}
\label{deltamb}
{{\delta m_b}\over {m_b^{\rm tree}}}=\tan \beta\left ({{2\alpha_s}\over 3\pi}
{{\mu m_{\tilde g}}\over m_1^2}
+{{h_t^2}\over {16\pi^2}}{{\mu A_t}\over m_2^2}\right ) ,
\end{equation}
where $m_{\tilde g}$ and $A_t$ are respectively the gluino mass and
the stop trilinear coupling, while  $m_1$ and $m_2$ represent 
effective SUSY masses out of the loop integrals. In particular $m_1$
roughly corresponds to the maximum between the sbottom and gluino
masses, while $m_2$ represents that between stop and higgsino (more 
exact expressions are found in \cite{HRS}). When the SUSY parameters
are all of the same order of magnitude, the typical size of the expression
inside brackets in eq.~(\ref{deltamb}) is $\sim 1\%$. 
Then for $\tan \beta={\cal O}(1)$ we can safely neglect them. On the
other hand for $\tan \beta \sim 10$ their effect could be a 10$\%$ reduction of
$R$ (notice that the sign of eq.~(\ref{deltamb})
is not fixed). As already stated, a $10\%$ reduction in
$R$ is critical for the bounds on $M_N$,
so that $\Omega_\nu \sim 0.1$ could be allowed for $\tan \beta > 10$. 
Notice however that $\delta m_b$ depends rather strongly on the
features of the SUSY spectrum and becomes negligible when either
$\mu$ or the gluino mass (and $A_t$) are somewhat smaller than the
squark masses. Only after an experimental
determination of the SUSY mass parameters,
will we be able to make conclusive statements on the $\tan \beta\gsim
10$ region. It remains however true that in a relevant
region of parameter
space (with small $\delta m_b$), the bounds shown in the figure are
valid.\footnote{
It must be added that the approximate symmetries which render large
$\tan\beta$ natural suppress $\delta m_b$ \cite{HRS}. On the other hand,
these approximate symmetries require $m_{SUSY}^2 \gsim\tan \beta m_Z^2$
because  of the LEP
constraint $\mu, m_{1/2} \gsim m_Z$. Therefore  the case $\mu
m_{\tilde{g}}/m_{SUSY}^2 \sim 1$ and the case
$\mu m_{\tilde{g}}/m_{SUSY}^2 \sim 1/\tan\beta$ require a comparable
fine-tuning in the Higgs sector \cite{NR,RSH}. However,
phenomenological constraints from the observed rate for $b \rightarrow s
\gamma$ again seem to favor the small $\delta m_b$ case \cite{RSH,RS}.
}

\item Corrections from physics above the GUT-scale. In our analysis 
we allowed a relatively large $h_t (M_G) < 2$, which implies
that the Yukawa Landau pole, and presumably some new physics,
can be as close as $8 M_G$.\footnote{If one requires perturbativity 
up to $M_{\it Planck}/\sqrt{8\pi}$, one needs $h_t (M_G) \leq 1.5$. 
On the other hand, notice that the value $h_t (M_G) = 3.3$ 
allowed in previous studies \cite{BBO} implies that the Landau 
pole is closer than $2 M_G$.}
In general, the presence of new physics at a scale 
$\Lambda > M_G$ induces non-renormalizable operators in the
effective theory at the GUT-scale. Such operators 
will in general affect eq.~(\ref{boundary})
with contributions scaling as powers of $M_G/\Lambda \equiv \eta$,
and one could expect $\eta \sim 0.1$ ({\it e.g.} from the above
observation).
Indeed, the order of magnitude of this ratio  naturally suggests that
flavour mixings and mass ratios are generated by the same type of
effects. From this viewpoint, it is legitimate to make the order of
magnitude interpretation $\eta \sim V_{cb} \simeq 0.05$.
Flavour scenarios of this type have been  depicted in
Ref.~\cite{ADHRS,KS}.
We will therefore use the indicative value above to
estimate the effects of non-renormalizable operators, which
we classify below.

A first class of operators is obtained by inserting $SO(10)$
breaking Higgs fields in the renormalizable term ${\bf 16_3 10_H
16_3}$. Our assumption that there be {\it small} irreps only
(${\bf 45}$'s, ${\bf 54}$'s, ${\bf 16}$'s, $\overline{\bf 16}$'s and
${\bf 10}$'s ) constrains the lowest possible correction to be ${\cal
O}(\eta^2) \sim 1\%$. These might come from
${\bf 16_3 10_H \langle 45 \rangle
\langle 45 \rangle 16_3}/\Lambda^2$ and ${\bf 16_3 10_H \langle 16_H
\rangle \langle 16_H \rangle 16_3}/\Lambda^2$.\footnote{Note that $\langle{\bf
54}\rangle$ preserves $O(6)\times O(4)$ (Pati-Salam group), 
and  cannot modify
eq.~(1). Also, ${\bf 16_3 10_H \langle 45 \rangle 16_3}$
respects eq.~(1) due to the symmetry between two ${\bf 16_3}$.}
Such effects can be safely neglected.

A second class of operators involves the other Higgs fields
${\bf 10'_H}$, ${\bf 16_H}$ and ${\bf \overline{16}_H}$, rather than
the ${\bf10_H}$ appearing in the renormalizable term. 
The only possible dimension five operators  are
${\bf 16_3 \bf 10'_H \langle 54 \rangle 16_3}$,
${\bf 16_3 10'_H \langle 45 \rangle 16_3}$,  ${\bf 16_3 16_H
\langle 16_H \rangle 16_3}$ and ${\bf 16_3 \overline{16}_H
\langle \overline{16}_H \rangle 16_3}$. None of them affects
$ h_b =  h_\tau $, while the last one may induce a correction 
${\cal O}(\eta)\sim 10\%$
to $h_t=h_N$. Such a correction does not affect our analysis, as 
we already stated (see discussion below eq.~(17)).
On the other hand, dimension six operators can affect $h_b=h_\tau$.
Actually, the relevance of this class of operators strongly depends
on the magnitude of $h_b, h_{\tau}$ themselves, {\it i.e.} on $\tan\beta$.
When $h_{b,\tau}\gsim
\eta$ (or $\tan\beta\gsim\eta m_t/m_b\sim 10$), dimension six
terms can give corrections which are ${\cal O}(\eta
^2/h_{b,\tau})\lsim \eta\sim$ 5--10\%. We recall that, for
$h_{b,\tau}\ll 1$, $H_d$ sits mainly in other Higgs fields than ${\bf10_H}$.
In fact, for $h_{b,\tau}\sim \eta$,
the above dimension five operators  become a natural {\it source} for these
Yukawa couplings. When $h_{b,\tau}\lsim \eta$, {\it i.e.}\/
$\tan\beta\lsim 10$, the 
dimension five operators themselves have to be suppressed, since they would
typically 
yield $ {\cal O}(\eta)$ Yukawa couplings. This could result from a flavor
symmetry. In this situation the dimension six operators become
relevant and can finally even dominate 
$h_{b,\tau}$. 
Since these
terms can give both $h_b=h_\tau$ and $3 h_b=-h_\tau$ (via
``composite'' $\bf 126$ combinations), or a mixture of both, the only
way not to lose predictivity completely is to suppose that only the first
type of terms exists. With this assumption the corrections to
$h_b=h_\tau$ are again expected to be at most ${\cal O }(\eta)$, 
and could be actually ${\cal O}(\eta^n)$ (with $n>1$) in more specific 
flavour models.

In short, $h_b = h_\tau$ is an automatic consequence of gauge symmetry
and field content alone for $h_{b,\tau} \gsim \eta$. For smaller
$h_{b,\tau}$, the relation $h_b=h_\tau$ should probably result from
flavor symmetries.

We are aware that the last class of corrections is less under our
control than all the other ones discussed so far. 
In fact, an opposite viewpoint would emerge if 
neutrino-oscillation experiments like CHORUS or NOMAD should find
$m_{\nu_\tau}\gsim $ eV. This discovery could indeed be used
to probe the existence of higher dimensional effects. 
For instance, should the observations suggest $M_N = 10^{12}$~GeV, then we
would need $R(M_G)=1.05$, 1.13, 1.23,  for $\alpha_s(m_Z)=0.110$, 0.117, 0.125,
respectively. 
It
is manifest that for the larger values of $\alpha_s$ we  need the second
class of effects
(and larger than expected), while for the smaller $\alpha_s$ GUT
thresholds alone could
account for the deviation of $R(M_G)$ from unity.
These remarks could be a useful guide in building realistic GUT
models. 

\end{enumerate}

Therefore, there are possible corrections to the prediction of $R(m_Z)$
for both $\tan \beta \gsim 10$ and $\tan \beta \lsim 10$ from very
different origins (indeed the reason is the same; namely $m_b \ll m_t$).
For large $\tan \beta$, non-logarithmic SUSY
threshold effects can modify the prediction, but they can be calculated after
the SUSY spectrum will be known. For small $\tan \beta$, corrections
arising from physics beyond the GUT-scale could lead to two
different classes of effects.
When only first class effects are present, like in 
Ref.~\cite{ADHRS}, then the  condition in eq.~(\ref{boundary}) is
robust and so are our bounds on $M_N$. On the other hand, when
$\tan\beta\ll m_t/m_b$, reasonable flavor physics scenarios
with the second class of operators may generically lead to
a ${\cal O}(5$--$10\%)$ shift in the boundary condition, which is
typically  non-negligible for our purposes. However, in specific models
this effect may also be further suppressed by symmetries.
Moreover, even with ${\cal O}(5$--$10\%)$ corrections, there still are 
stringent limits on $M_N$ for $\alpha_s (m_Z) \gsim 0.12$.
Note also the new physics scale $\Lambda$ could be larger $\gsim 100 M_G$
when $h_t (M_G) \lsim 1.5$.
At large $\tan\beta$, GUT-scale uncertainties are
under control, so that
the measurement of SUSY parameters will
be enough to make our analysis very accurate.

{\bf 7.} In summary, we have studied the impact on the low energy value
of $m_b/m_\tau$ 
of an intermediate mass right-handed neutrino in a class of $SO(10)$
models. The analysis generally results in a lower bound on the mass $M_N$ of
the right handed neutrino, and thus in an upper bound on the $\tau$ 
neutrino mass and on its contribution $\Omega_\nu$ to the hot dark
matter density in the present universe. 
In order to do so we have performed a two-loop study of the RGE and
discussed threshold effects  at the GUT and SUSY scales.
Depending on the values of $\alpha_s$ and $\tan\beta$
the bound varies from a  very strong one  to a weaker one.

{\it i)} When $\tan\beta\gsim 10$, there are uncertainties in the 
bound on $M_N$ from the yet-unknown non-logarithmic SUSY
threshold corrections. When these effects are maximal, 
then $M_N$ is allowed to span all the phenomenologically interesting region
(down to $M_N\sim 10^{12}$ GeV). On the other hand, when these effects
are small, then only for very large $\tan\beta\gsim 50$ can we reach
the cosmologically interesting region 
$M_N\lsim 10^{13}$ GeV. We stress that effects at the SUSY-scale
threshold will be known once SUSY particles are discovered and their
parameters measured, so that such effects will be no longer ambiguities.
Instead, we will be able to evaluate the corrections, which might even
make the constraint stronger.

{\it ii)} For small $\tan\beta$, we get rather stringent bounds on
$\Omega_\nu$. For 
$\alpha_s(m_Z)=0.11,0.117,0.125$ we respectively get $\Omega_\nu h^2 <
4\times 10^{-3}, 3\times 10^{-5}, 6\times 10^{-6}$ for $\tan \beta
\leq 10$. 
Corrections from SUSY thresholds do not modify these bounds.
If we allow for ``maximal'' GUT
threshold 
effects of ${\cal O}(5\%)$, these bounds can go up by 
about two orders of magnitude. We point out that there might be additional
corrections to $h_b / h_\tau = 1$ at $M_G$ from physics beyond $M_G$.
Assuming that the new physics is responsible for the flavor structure,
we estimate the size of the corrections to be ${\cal O}(V_{cb}) \sim 5\%$.
Even taking all these possible ambiguities into account, the
cosmologically interesting region $\Omega_\nu \gsim 0.1$ is allowed
typically only for $\alpha_s (m_Z) \lsim 0.12$. 
Improvement in the experimental knowledge of
$\alpha_s$ \cite{Matsui} 
will allow tighter bounds, especially if the value will converge towards
the present central one $\gsim 0.12$.

\section*{Acknowledgements}
This work was supported by the United States Department of Energy
under contract DE-AC03-76SF00098 (A.B. and H.M.), and by NSF grants
PHY-91-21039 (R.R.), PHY-89-04035 (H.M. and R.R.). The work of A.B. is
also sponsored by an INFN
fellowship. H.M. and R.R. would like to acknowledge the hospitality
and support of the Institute for Theoretical Physics at Santa Barbara,
where this work was started. They also wish to thank the organizers 
and the participants of the Weak Interaction Program at ITP (spring '94) for  
the friendly atmosphere and the many stimulating discussions.

\section*{Note Added}

After completing the main part of the work described in this letter 
and presenting it at a conference
\cite{SUSY94}, we received a preprint by Vissani and Smirnov 
discussing a similar topic 
\cite{Smirnov}. Their basic conclusion is that an intermediate mass
right-handed neutrino disfavors $b$--$\tau$ Yukawa unification, or
{\it viceversa},\/ which is the same as ours.
However when deriving quantitative lower bounds on $M_N$,
their analysis is based on one-loop RGE with tree-level
matching, while ours is based on two-loop RGE with one-loop matching. 
We consider this to be necessary due to the high sensitivity of the
bounds on $M_N$ to ${\cal O}(5$--$10\%)$ effects
which might arise from
SUSY-scale and GUT-scale
threshold corrections. 
In addition, they do not discuss the possible
relevance of points (1) and (2) we discussed above. However, in the
cases where all such effects are negligible, our result is consistent
with theirs.

\newpage
\section*{Figure Captions}
\renewcommand{\labelenumi}{Fig.~\arabic{enumi}}
\begin{enumerate}
\item An illustration of the effect of the right-handed neutrino on
$R(m_Z) = m_b(m_Z)/m_\tau(m_Z)$. The curves show the dependence of $R(m_Z)$
on $h_t (M_G) 
= h_N (M_G)$, for the values of $M_N = M_G$, $10^{15}$~GeV,
$10^{14}$~GeV, $10^{13}$~GeV, $10^{12}$~GeV, and $10^{10}$~GeV. The
experimentally allowed values of 
$R(m_Z)$ are also shown. Other input parameters are taken as 
$m_{SUSY} = m_A = m_Z$, $h_b (M_G) = h_\tau
(M_G) = 0.01$, and $\alpha_s(m_Z) = 0.11$ (a) or 0.12 (b).
\item Lower bounds on the right-handed neutrino mass $M_N$ from
$b$--$\tau$ Yukawa unification. Curves for three values of $\alpha_s =
0.110$ (solid), 0.117 (dots) and 0.125 (dash) are shown, both for the
SUSY particle spectra ({\it a}) $m_{SUSY} = 1$~TeV, $m_A = m_Z$ (lower) and
({\it b}) $m_{SUSY}= m_A = m_Z$ (upper); see text. The possible cosmic energy
density $\Omega_\nu h^2$ of the $\tau$-neutrino hot dark matter is also
shown for comparison in dot-dashed lines.
\end{enumerate}

\end{document}